\documentclass[12pt,preprint]{aastex}

\newcommand{\cwsdeg}{^{o}}

\shorttitle{Preliminary PanSTARRS Throughput Determination}
\shortauthors{C.W. Stubbs et al}
\begin{document}
\title{Precise Throughput Determination of the PanSTARRS Telescope and the Gigapixel
Imager using a Calibrated Silicon Photodiode and a Tunable Laser: Initial Results}
\author{Christopher W.~Stubbs, Peter Doherty, Claire Cramer,\\ 
Gautham Narayan, Yorke J.~Brown} 

\affil{Department of Physics, 17 Oxford Street \\
Harvard University, Cambridge MA USA}
\email{stubbs@physics.harvard.edu}

% doherty@physics.harvard.edu, ccramer@cfa.harvard.edu, narayan@physics.harvard.edu, yjb@yorkebrown.com, ned.henry@gmail.com 

\author{Keith R.~Lykke, John T.~Woodward}
\affil{National Institute of Standards and Technology, \\
100 Bureau Drive, Gaithersburg MD 20899}

%\email{keith.lykke@nist.gov, steven.brown@nist.gov, john.woodward@nist.gov}

\author{John L.~Tonry}
\affil{Institute for Astronomy, University of Hawaii, 2680 Woodlawn Drive, Honolulu HI 96822}

%\email{jt@ifa.hawaii.edu}

\begin{abstract}

We have used a precision calibrated photodiode as the fundamental metrology reference 
in order to determine the relative throughput of the PanSTARRS telescope and the 
Gigapixel imager, from 400 nm to 1050 nm. 
Our technique uses a tunable laser as a source of 
illumination on a transmissive flat-field screen. 
We determine the full-aperture system throughput as 
a function of wavelength, including (in a single integral measurement) the mirror reflectivity, the transmission functions of the filters and the corrector optics, and the detector quantum efficiency, by comparing the light seen by each pixel in the CCD array to that measured by a precision-calibrated silicon photodiode. This method allows us to determine the {\it relative} throughput of the entire system 
as a function of wavelength, for each pixel in the instrument, without observations of celestial standards. We present promising initial results from this characterization of the PanSTARRS system, and we use synthetic photometry to assess the photometric perturbations due to throughput variation across the field of view. 
  
\end{abstract}


\vfil
\eject
\clearpage

\section{INTRODUCTION}
\label{sec:intro}

Multiband ground-based photometry plays a central role in a variety of forefront topics in astrophysics. Examples
include mapping the expansion history of the Universe with type Ia supernovae, determining redshifts to galaxies and clusters using photometric redshifts, 
monitoring the spectral evolution of gamma ray bursts, and detecting and characterizing extrasolar planets with transits. 

A specific contemporary scientific objective that requires a clear understanding
of system sensitivity across filter bands is using type Ia supernovae to probe the 
history of cosmic expansion. Imagine that the spectrum of supernova emission 
were concentrated in a narrow wavelength range, $\delta \lambda$. Observing 
supernovae at increasing redshift would move the center of emission into redder
passbands. In order to precisely compare the luminosity distance vs.~redshift, we
must clearly have a careful calibration of the {\it relative} sensitivity across
the passbands of interest, with a well-understood metrology foundation. 
Ascertaining the equation of state parameter $w=P/\rho$ of the dark energy, 
and searching for any variation with redshift, will require supernova flux measurements 
with percent or better {\it precision} ({\it e.g.}~\cite{w}).  The SNLS 
collaboration has assessed photometric calibration as the dominant source
of systematic uncertainty in their program to measure the equation of state
parameter $w$ of the dark energy \citep{SNLS09}. 

Another opportunity for exploiting high accuracy 
astronomical flux measurements will arise with the launch of the Gaia astrometric
satellite mission, which should provide $\sim10^5$ stars with distances accurate
to better that 0.1\% \citep{gaia}. Providing a commensurate {\it accuracy} for 
photometric measurements will allow for detailed comparisons between observed
stellar luminosities and model atmospheres.  

These ground-based photometric measurements face a number of calibration challenges. One is knowing the instrumental sensitivity as a function of wavelength. 
Another is accounting for absorption and scattering in the Earth's atmosphere. 
The traditional methodology of photometric calibration uses some combination of  (1) observations of standard sources at the same airmass as the program objects, or (2) measurements of extinction coefficients in the passbands of interest. 
In this approach, the spectra of celestial sources 
constitute (often implicitly) the  
standard for the measurement. These spectrophotometric standards are either
based on ground-based blackbody measurements made decades ago \citep{Hayes75,meg95}, or in the case of DA white dwarfs, on theoretical model atmospheres ({\it e.g.} \cite{Holberg06} ).
Since broadband CCD photometry 
of celestial sources takes an integral of photon flux across a 
broad optical passband ($u,g,r,i,z$ or $y$, for example), 
it is an ill-posed problem to infer the system sensitivity function vs. wavelength
from broadband photometric data alone, especially 
through a variable atmosphere. 

The determination of the effective system throughput
has often been done in a piecemeal fashion, taking benchtop measurements of filter transmission, optical catalog values for reflectivity and 
glass transmissions, and vendor descriptions of typical detector quantum efficiency
 (QE). 
The cumulative systematic errors in this procedure limit the precision that can 
be achieved, and this approach lacks resolution at the pixel scale. 
Attempting to adjust passbands determined in this fashion to 
match observations to synthetic photometry is ambiguous: should the
passbands be broadened, shifted up or down in wavelength, or should a 
grey multiplicative scaling be applied? 

In order to address these concerns, and to push towards an improvement in 
photometric precision and accuracy, \cite{Calib06} advocated breaking the 
spectrophotometric calibration problem into two distinct 
measurements: 
1) the determination of instrumental sensitivity, using laboratory-calibrated
detectors as the fundamental metrology standard, and 2) measuring directly
the optical transmission of the atmosphere. 
We will focus here on our implementation of a technique for
measuring, {\it in situ},  the relative throughput of the entire PanSTARRS apparatus, 
relative to a calibrated photodiode. 

We can forgo the use of a celestial calibration {\it source} in favor of a well calibrated 
{\it detector} as the fundamental metrology reference for astronomical photometry. 
National Institute of Standards 
and Technology (NIST) has calibrated Silicon photodiode quantum efficiencies 
at the 10$^{-3}$ level. This is an order of magnitude more precise than any celestial spectrophotometric source, 
either empirical or theoretical. In fact, the primary SI metrology  reference for electromagnetic flux is now detector-based rather than source-based \citep{NIST}. 
There is ongoing progress at NIST in extending both the wavelength range and
accuracy of detector calibration for metrology applications. 

We assert that no celestial spectrophotometric source currently offers a photon spectral 
distribution that is known at the $10^{-3}$ level, and the prospect of achieving this 
precision is one of the main attractions of the method described here.  

This paper presents promising initial results from  
an integrated measurement of the total system throughput. We use 
full-pupil illumination with monochromatic calibration light to measure the entire 
optical train of the apparatus, including the mirror, corrector optics, filter and detector.
The spatial resolution is at the pixel scale, with 1 nm  spectral resolution.  
An earlier realization of this technique, with the 
Mosaic imager on the 4 meter Blanco telescope at CTIO, was described in 
\cite{CTIOpaper}. 

The determination of and correction for variation in atmospheric transmission is a 
significant challenge. A companion paper \citep{Burke10} describes
progress in the precise determination of atmospheric transmission.
We also refer the interested reader to \cite{Calib06} and \cite{Atmos} for a 
discussion of this issue.  The Pierre Auger collaboration has implemented
\citep{Auger10}
a comprehensive suite of atmospheric monitoring instruments for the calibration 
of optical transmission in atmospheric fluorescence detection of high energy
cosmic rays, and we advocate taking a similar approach for optical and infrared
astronomical observations. 

The merits of achieving improved photometric accuracy are spelled out in 
\cite{Kent09}, and there is considerable work under way to achieve 
improved accuracy and precision. 
Many of these are described in \cite{ASP}. The ACCESS project \citep{ACCESS} 
plans to conduct 
precise spectrophotometry from a sounding rocket. \cite{Bohlin07} 
describes spectrophotometric measurements at the 1\% level using HST. 
Adelman and colleagues \citep{Adelman07} are pursuing ground-based
precision spectrophotometry. 
\cite{SLR} describe the use of the stellar locus in color-color space 
for precise color determinations. 
\cite{Albert09} discussed the merits of
a satellite-based calibration source, which would provide an 
opportunity to determine both atmospheric transmission and apparatus throughput 
in the same measurement, providing the satellite could be accurately 
tracked by the telescope.

The Sloan Digital Sky Survey has used monochromatic light to measure and 
monitor the spectral sensitivity function of the SDSS camera when removed
from the telescope \citep{SDSS10}. Our approach uses full aperture illumination
of the entire optical system rather than measuring only the camera's response. 

We are not currently attempting to establish 
an absolute calibration of the PanSTARRS system, 
in units of photons cm$^{-2}$ s$^{-1}$ nm$^{-1}$, but rather to establish the {\it relative} system
throughput across wavelength, in arbitrary units. A specific example of 
the importance of understanding the relative photometric zeropoints across
filters is when type Ia supernovae are used to map out the history of cosmic 
expansion. A complete understanding of the effective system passband, and the 
associated zeropoints, is essential. Type Ia supernova cosmology is one of the 
many science objectives planned for the PanSTARRS survey. A single measurement
of absolute detection efficiency, at one wavelength, would suffice to place
the data presented here onto an absolute flux scale. An approach to full-aperture
determination of absolute detection efficiency is described by the Auger collaboration
in \cite{Auger04}.

The calibration technique described here can be used in conjunction with 
more traditional photometric calibration methods. Using a diversity of calibration methodologies provides an opportunity
to assess consistency and to quantify potential sources of systematic error. 
This approach is not limited to imaging instruments. Spectrophotometric 
sensitivity functions for dispersive instruments can be acquired in a 
similar fashion. 

A specific motivation for our calibration program is the requirement of knowing the system 
throughput well enough to perform high confidence color determination for the 
PanSTARRS survey. As described below, this technique also allows us to 
monitor filter transmission curves (to check for variation due to hygroscopic 
effects, for example), and any other changes in instrumental performance.

This paper first presents an overview of the experimental method and apparatus
in Section \ref{sec:methods}, and the observations we obtained in Section 
\ref{sec:observations}. Data processing is outlined in Section \ref{sec:processing}, 
and results are shown in Section \ref{sec:results}. Section \ref{sec:systematics} 
presents a preliminary assessment of potential sources of systematic error, 
followed by our conclusions in Section \ref{sec:discussion}. 

\section{EXPERIMENTAL METHOD AND APPARATUS}
\label{sec:methods}

We have implemented the calibration philosophy outlined in \cite{Calib06}, 
which exploits the availability of well-calibrated detectors to characterize 
astronomical apparatus. The reader
is strongly encouraged to consult that paper for a full description of the 
motivation of this approach, and the related formalism. 

\begin{figure}[htbp]
\begin{center}
\centerline{\includegraphics[width=5.5in]{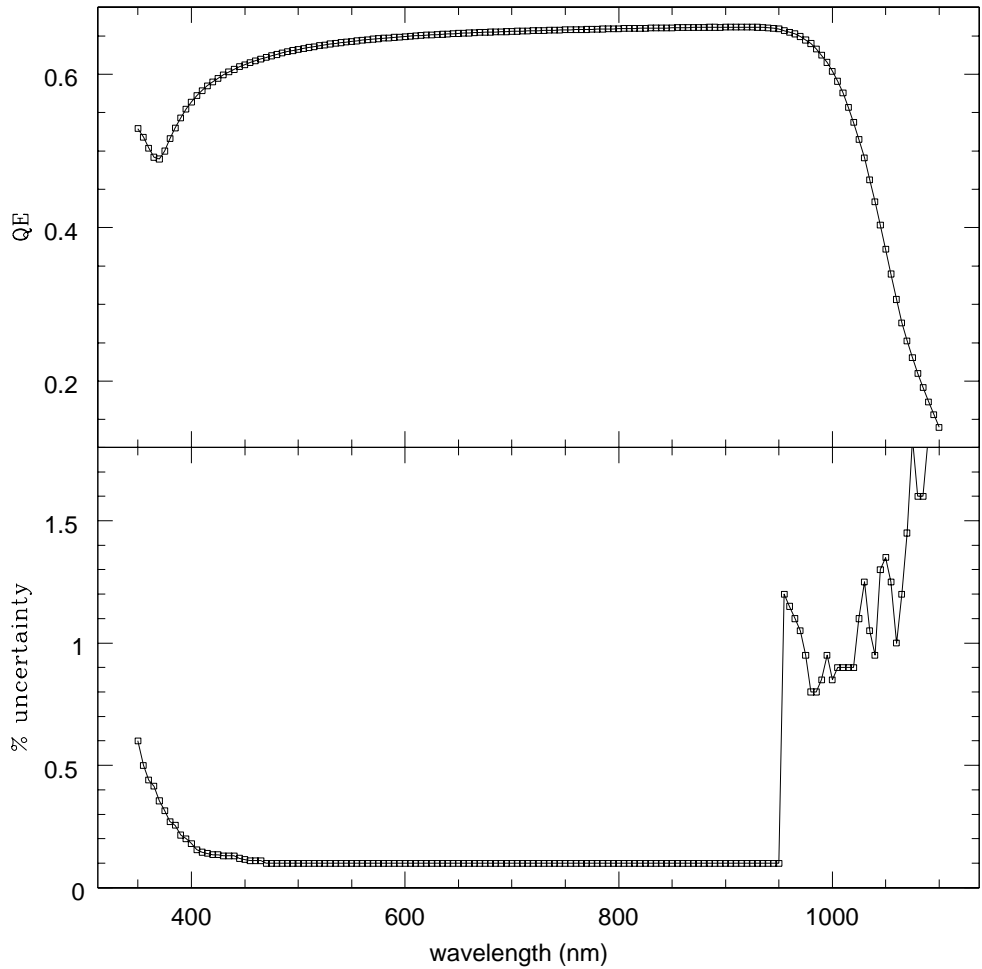}}
\caption{Quantum Efficiency Curve and Fractional Uncertainty for NIST-calibrated Photodiode. The upper plot shows the photon detection 
efficiency vs. wavelength for a Hamamatsu 2281 photodiode.
This calibrated response curve 
is the standard against which we measure system throughput, and is better determined than any celestial 
spectrophotometric source. The lower panel shows that over the majority of the wavelengths used
for CCD imaging, for the photodiode we used the fractional calibration uncertainty in QE is 0.1\%. Only in the $y$ band, with $\lambda>950$~nm, 
and for $\lambda < 400$~nm (which PanSTARRS does not use) is this photodiode's calibration poorer than a part per thousand. The discontinuous jump in 
calibration uncertainty
for $\lambda>$950 nm arises from a change in the metrology comparison 
method used by NIST in 2005, when our reference diode was calibrated. More recent
work at NIST \citep{NIRcalib09} extends high precision calibration out to 
1.6 $\mu$m, with the prospect of achieving 0.01\% accuracy.}
\label{fig:NIST}
\end{center}
\end{figure}

Our system uses the smooth and well-characterized detection efficiency of a 
NIST-calibrated
photodiode as the standard against which we calibrate the apparatus.   
The main principle of this technique is to establish the QE($\lambda$) curve of the 
calibration photodiode as the metrology foundation for the relative throughput 
measurement. Figure \ref{fig:NIST} shows the smooth dependence of the photodiode
QE vs. wavelength. 

\begin{figure}[htbp]
\begin{center}
\centerline{\includegraphics[width=6in]{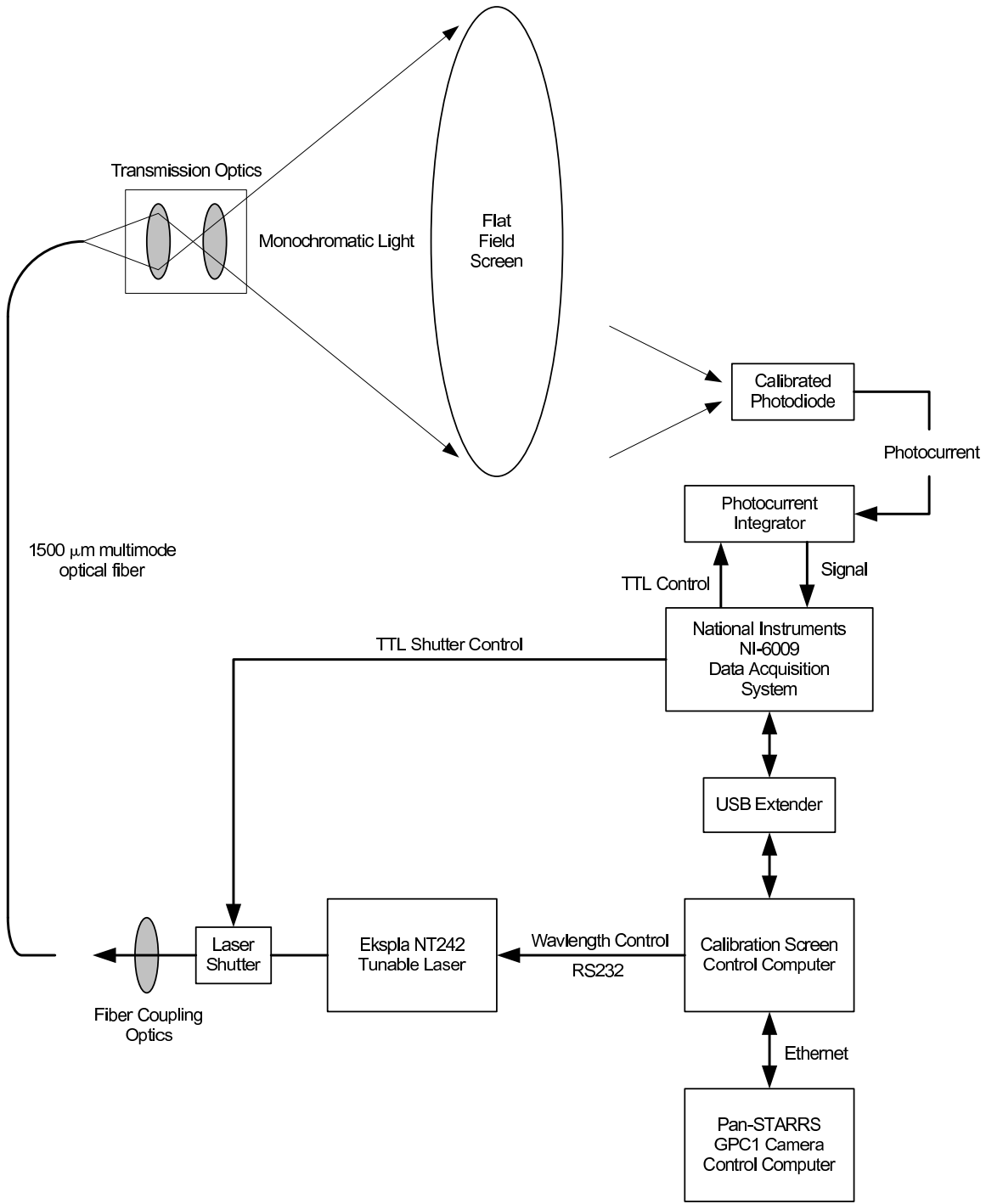}}
\caption{Schematic diagram of calibration system configuration. Monochromatic light from the tunable laser is projected onto the full-aperture flat-field screen. The calibrated photodiode is used to 
monitor the total laser light delivered to the input pupil.}
\label{fig:system_diagram} 
\end{center}
\end{figure}

An overall conceptual diagram of the arrangement of the apparatus is shown in Figure \ref{fig:system_diagram}.  
We project light from a tunable laser onto the flat-field screen in the dome. This 
screen provides full-aperture illumination of the PanSTARRS telescope. We measure the 
flux from the screen with a calibrated photodiode. We
then compare the flux detected by each pixel in the instrument to that detected by the calibrated photodiode. 

Performing this measurement at a succession of discrete wavelengths allows us to determine full system throughput as a function of wavelength. 
By interspersing calibration images with equivalent exposure
times taken with the laser off, we can compensate for flux from ambient light in the dome. 

\subsection{The PanSTARRS Telescope and the Gigapixel Imager}

The PanSTARRS system is a high etendue wide-field imaging system, designed
for dedicated survey observations. The system is installed on the peak of 
Haleakala on the island of Maui in the Hawaiian island chain. Routine 
observations are conducted remotely, from the Waiakoa Laboratory
in Pukalani.  We provide below a terse summary of the PanSTARRS survey
instrumentation. Further details can be obtained from the project's web 
site.\footnote{\tt http://pan-starrs.ifa.hawaii.edu/public/project-status/project-status.html}

\subsubsection{The PanSTARRS Survey Telescope}

The PanSTARRS optical train uses a 1.8~meter diameter $f$/4.4 primary
mirror, and a 0.9~m secondary. The resulting converging beam then
passes through two refractive correctors, a $48$~cm~$\times~48$~cm
interference filter, and a final refractive corrector that is the
dewar window.

\subsubsection{The PanSTARRS Gigapixel Camera}

The PanSTARRS imager \citep{GPC} comprises a total of 60
$4800\times4800$ pixel detectors, with 10~$\mu$m pixels that subtend
0.258~arcsec.  The total field of view of the system is 3.3
degrees. The detectors are back-illuminated CCDs manufactured by
Lincoln Laboratory.  The detectors are read out using a StarGrasp CCD
controller \citep{StarGrasp}, with a readout time of 7 seconds for a
full unbinned image.

The CCDs in the PanSTARRS camera are deep-depletion devices, 75~$\mu$m
thick, which increases the QE in the region around 1~$\mu$m and
minimizes the effect of fringing from sky lines.  The instrument has a
filter changing mechanism that contains 6 large-format interference
filters, {\it g,r,i,z,y} and {\it w}.

The temperature of the CCD array was -75C, for the data presented
here.  For the $y$ band, where the red edge of the passband is
partially defined by the falling silicon QE curve, the detector QE is
temperature sensitive.  We made explicit note of the temperature
registered by the StarGrasp controller for each flat-field image
obtained.  In the future we expect to measure the temperature
dependence of the system efficiency, arising in the detectors because of
this phonon-assisted photoconversion regime in silicon.

The shutter for the Gigapixel array is a dual-blade linear curtain shutter, 
that provides millisecond-accuracy timing information on actual shutter open 
times. This accuracy and repeatability is important for proper
background subtraction if ambient light correction is substantial. 
Our method ensures that the shutter is fully open 
during the interval over which we inject laser light, so even instruments with iris-type
shutters could be calibrated with minimal shutter artifacts. 

\subsection{Monochromatic Calibration Apparatus}

Light from the tunable laser was transported through an optical fiber 
to a back-illuminated flat-field screen. A photodiode with a transimpedance 
amplifier was used as a coarse light level monitor, and was useful for 
adjusting the light intensity to the desired level. The calibrated photodiode
monitored the light emanating from a portion of the screen, and was 
fed into an integrator circuit as described in more detail below. 
A shutter was used to interrupt the light into the source fiber. 

An observing sequence entailed the following stages:

\begin{enumerate}

\item{} Set desired wavelength,		

\item{} Open camera shutter,	

\item{} Wait 1 sec,

\item{} Open laser shutter,

\item{} When time or flux threshold satisfied, close laser shutter,

\item{} Wait 1 sec to ensure full light dose received by camera, and to obtain post-laser ambient light level in photodiode signal, 

\item{} Close camera shutter and read out image

\item{} Obtain another image and photodiode series, for same camera shutter open 
time, with laser off, for ambient light compensation.

\end{enumerate}

The tunable laser was located on level 1 of the PanSTARRS dome. A
multimode optical fiber, NA=0.3, with low OH content, with a diameter of 1.5~mm, and clad in semirigid 
conduit (Ceramoptic part number SMA2/WF1500/1590T30/SS/20.0M)
\footnote{
Certain trade names and company products are mentioned in the text or identified in an illustration in order to adequately specify the experimental procedure and equipment used. In no case does such identification imply recommendation or endorsement by the National Institute of Standards and Technology, nor does it imply that the products are necessarily the best available for the purpose.}  carried the laser light up
through the telescope structure and onto an optical bench assembly that was mounted
above the flat field screen. 

\begin{figure}[htbp]
\begin{center}
\epsscale{0.6}
\plotone{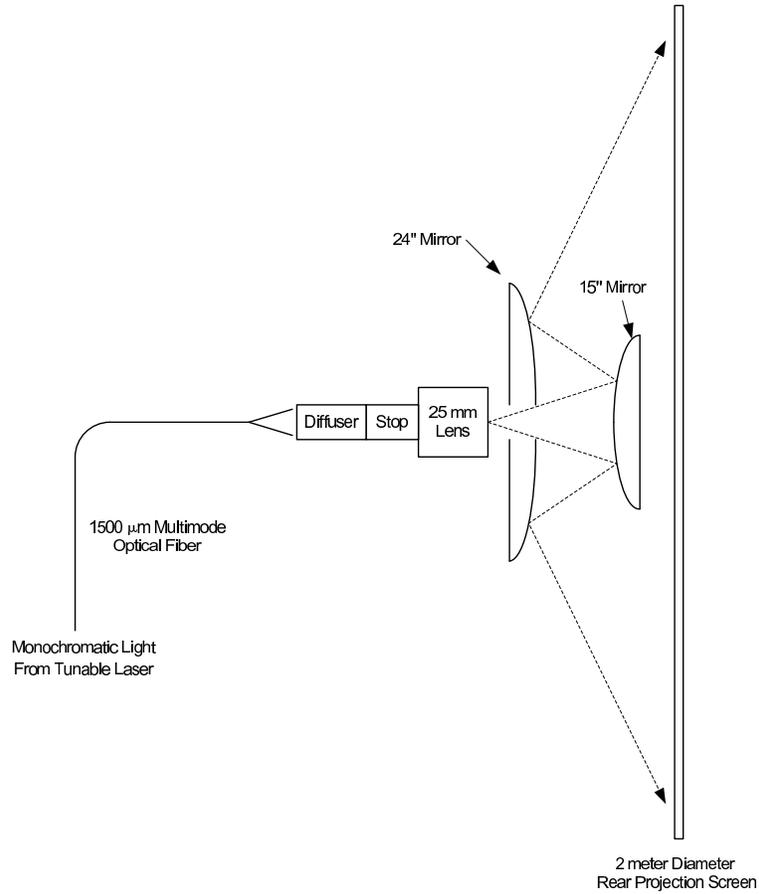}
\caption{Calibration light from the optical fiber is collimated with an achromatic
25mm focal length lens, followed by an engineered diffuser. An aperture stop
after the diffuser is imaged onto a backlit flatfield screen using a combination 
of a broadband lens and a pair of convex acrylic mirrors. The projection system
produces an image of the aperture on the back-illuminated 
transmissive flatfield screen.  The calibration diode monitors light 
that emanates from the screen.}
\label{fig:launchoptics}
\end{center}
\end{figure}

One challenge posed
by the PanSTARRS system is that the space available in the dome is minimal. To achieve a full-pupil illumination of the telescope, we constructed a projector box that is
mounted on the inner wall of the enclosure. This illuminates a transmissive 
flat-field screen (obtained from the Day-lite Corporation) that sends calibration 
light into the front of the telescope. 
We illuminated the rear of the flat-field screen by 
sending light that emerged from the
fiber through the optical layout shown in Figure \ref{fig:launchoptics}.  

The PanSTARRS enclosure and telescope cannot perform independent
rotations in azimuth. The telescope is always in the plane of the enclosure slit. The calibration screen and projector box are therefore mounted ``behind'' the 
enclosure slit, and the elevation axis of the telescope is rotated past vertical, to 
125 degrees from the horizon, to point at the screen. 

Each unvignetted element of area $dA$ on the flatfield screen illuminates every pixel of the imager, 
since the optical system maps angles at the pupil to position in the focal plane. 
Each pixel of the imager correspondingly receives light that emanates from the 
everywhere on the screen, at the 
angle of arrival that maps onto that pixel.  Different regions of the focal plane
receive flux from slightly different areas of the illumination screen, but as long as the 
screen surface brightness distribution is independent of wavelength we can 
reliably measure the relative spectral response of each pixel.  

As long as the illumination pattern
on the screen is independent of wavelength, at the 10\% level, the laser 
illumination allows us to determine (at the 1\% level or better) 
the wavelength dependence of the sensitivity of {\it each} pixel. 
We can correct for the nonuniform illumination in our laser dome flats in the same way
illumination corrections are traditionally used to correct dome flats. 

We have also installed a digital micromirror
array in the projection box, that can be fed with an optical fiber, and we intend
to use this to quantitatively assess the effect of screen illumination 
non-uniformities. Operating the 
calibration system in this fashion is the equivalent of projecting an image using 
a digital projector, except the light source can be either a white light source, or
monochromatic light from the tunable laser. This capability was not employed
for the results presented here, and so the entries in the systematic error budget
due to illumination non-uniformity contain estimates, for the time being. 

\subsection{Tunable Laser Source.}

We used an Ekspla NT242 OPO laser as a tunable source of monochromatic light. 
A 1 KHz pulsed Nd:YAG laser at 1.064 $\mu$m is the initial source
of photons. This light passes through a pair of non-linear optical crystals that upconvert the light to $\lambda$=355 nm. These UV photons are then run through an optical parametric oscillator (OPO) that splits each UV photon into a pair of photons, conserving both energy and momentum. 
The orientation of the OPO crystal relative to the incident beam can be adjusted so as to select a specific wavelength of interest. The wavelength of the output beam can thereby be tuned over a wavelength range of 
210~nm to 2.3~$\mu$m. 

From the photon pair produced, the upper or lower frequency light is selected by exploiting the fact that these two beams have orthogonal linear polarizations. 
There is a ``degeneracy point''  in the OPO system at 
$\lambda=$2~$\times$~355~nm =710~nm where the OPO power drops, but this 
did not pose a significant problem. 
For wavelengths $\lambda <$ 410 nm the Ekspla laser uses up-converted light
from OPO emission at 2$\lambda$, and both the output power and spectral purity
are considerably less favorable. We elected in this run  
not to extend our measurements below
410 nm, pending the installation of a short-pass filter in the beam. 
Although the light emitted from the Ekspla NT242 OPO is polarized, as the light is transmitted through the 
optical fiber its polarization becomes randomized.   
 
The 5 nsec Ekspla NT242 OPO pulses have the advantage that the coherence length of the 
laser light is correspondingly short, so the light on the flatfield screen shows no evidence
of speckle patterns that would afflict a continuous narrowband source. The 
spectral breadth of the tunable light is about 0.1 nm at 400 nm and 0.5 nm at 1 $\mu$m. 
The light intensity from the Ekspla NT242 OPO is somewhat adjustable. The Nd:YAG laser has a variable time delay between the flashlamp pulse and the Q-switch driven dump of the laser cavity. 
Changing this time delay varies the intensity of the light emitted by the Nd:YAG laser, 
and hence the intensity of the tuned light as well. 

\subsubsection{Wavelength Calibration}

We used a spectrograph from Ocean Optics (model Jaz) to monitor the wavelength of the light being sent to the flat-field screen. 
We used this spectrograph in conjunction with a Hg-Ar line lamp 
module (Ocean Optics HG-1) which constitutes our primary wavelength standard.
We tuned the Ekspla NT242 OPO to a succession of wavelengths that corresponded to 
lines emitted by the arc lamp source, and we compared the line centroids 
for the laser light to those from the arc lamps. The results are shown in Table
\ref{tab:lambda}. The table shows the 
line used, the Ekspla NT242 OPO setting, the reported centroids for the laser
and arc lamp, and the magnitude of the centroid difference. The Ekspla NT242 OPO 
wavelengths are accurate to $\sigma_{\lambda}\sim$~0.2~nm, 
comparable to the intrinsic line width of the laser. 
The wavelength calibration of the laser is more accurate than the 
wavelength solution of the spectrograph, but we stress that the spectrograph
was used only 
as a {\it differential} instrument, comparing the wavelengths of the arc lamp to that
emitted by the laser. 

\begin{table}[htdp]
\begin{center}
\caption{Wavelength Calibration.}
\begin{tabular}{ccccc}
\hline
\hline
Line & Laser setting & Laser centroid & Arc centroid & $|$Difference$|$ \\
(nm) & (nm) & (nm) & (nm)& (nm) \\
\hline
Hg 435.83 & 435.8 & 436.8 & 437.0 & 0.2\\
Hg 546.07 & 546.1 & 547.7 & 547.6 & 0.1 \\
Ar 696.54 & 696.5 & 698.1 & 698.0 & 0.1 \\
Ar 763.51 & 763.5 & 765.3 & 764.9 & 0.4 \\
Ar 800.62 & 800.6 & 802.2 & 802.4 & 0.2 \\
Ar 912.30 & 912.3 & 913.1 & 913.5 & 0.4\\
\hline
\end{tabular}
\end{center}
\label{tab:lambda}
\end{table}%

\subsection{Monitoring Light with Calibrated Photodiode}

We configured a NIST-calibrated silicon photodiode to monitor light emanating
from the flat-field screen. The photodiode is a Hamamatsu S2218 whose 
quantum efficiency was tied to our NIST-calibrated identical part, serial number 
I220. The calibration diode was equipped with a blackened ``snout'' tube that 
limited its angular field
of view. The snout tube had a length of  76.2 mm and an inner diameter of 25.4 mm. 
This limited the photodiode's field of regard to 18 degrees, which 
from the diode's location spanned
about a quarter of the surface of the emissive screen.  The temperature
of the calibration photodiode for the data presented here was typically around 
12$\cwsdeg$C. 
By avoiding 
any imaging optics we avoid introducing any unwanted wavelength dependence in the calibration photodiodes's signal chain. 

\subsection{Photodiode Electronics and Instrument Interface}

The photodiode fed an integrator circuit that we used to monitor the 
light emitted by the screen. We used a polypropylene 1000 pF 
capacitor (Panasonic type ECQP) in the integrator. We consider this 
preferable to a transimpedance amplifier, for a number of reasons. 
First, the 5 ns pulses from the Ekspla NT242 OPO produce photocurrent bursts from the
photodiode, and the finite gain-bandwidth product of a transimpedance configuration 
essentially attenuates the signal. On the other hand the integrator configuration 
assures that all the photoelectrons generated in the calibration photodiode eventually
appear across the integrating capacitor, with no loss of signal. Perhaps more 
importantly, we are primarily interested in the total laser light delivered to the 
system, and this favors the integrator circuit configuration.
The integrator output signal then ran through a unity gain buffer amplifier and
into the data acquisition system. This light-integration approach was also important in 
minimizing any deleterious effects due to pulse-to-pulse intensity variations in the Ekspla NT242 OPO output. 
The integrator was reset on command (with a mechanical relay, to avoid 
charge injection issues) using one of the digital outputs on the 
data acquisition module. 

The  data acquisition module was model NI USB-6009 from 
National Instruments, with 14 bit A/D conversion and 12 digital input/output bits which 
was connected to the data acquisition computer with a USB interface. 

We did consider potential systematic errors arising from the 5 nsec intense bursts of photons we
delivered to the focal plane. The photoconversion response of silicon in this regime has
been shown to be linear and well behaved (\cite{Stuik02}, \cite{Vest03}). This
applies to both the CCDs in the instrument as well as the calibration photodiode. 

\subsection{Data Collection Software and Architecture}

We adhered to the overall philosophy of the PanSTARRS system, by establishing
the calibration computer as a server that was accessible over the network in the dome. 
Programs requesting services (such as setting a laser wavelength, etc.) communicated
via socket connections to this server. The calibration server computer is a node that 
is responsible for interfacing with the tunable laser, with the photodiode data 
acquisition system, and the wavelength monitoring spectrograph. 

PanSTARRS collects multi-extenstion FITS images, and appropriate header
keywords were populated with sufficient calibration metadata so that the
images and their headers constitute a complete data set for downstream 
analysis. These header keywords include light dose information from the 
photodiode and integrator, and the wavelength of the calibration source for 
each image. 

Data collection is scriptable, allowing sequences of wavelengths to be collected
automatically. The calibration server allows for two modes of operation. In one, 
the laser shutter is opened for a fixed period of time. We prefer to use a second 
mode, where the laser light was kept on until the integrator signal reached a 
pre-determined threshold. The data presented here were all obtained in the 
fixed-threshold mode. If the monitor photodiode's quantum efficiency were 
wavelength independent, this would amount to a fixed dose of photons for each 
exposure. 

\section{OBSERVATIONS}
\label{sec:observations}

We obtained the data presented here on UT Dec 12 and 13, 2009. 
The majority, but not all,  of the observations were obtained at night
to minimize deleterious
effects of stray light in the dome. We determined that a spacing of 2 nm was
adequate to sample the structure in the wavelength-dependent system response, 
including both filter response function variations and fringing in the CCDs. 
We obtained a succession of laser-on and laser-off images, each with a corresponding
data set from the monitor photodiode. Scanning each filter took about an hour
of automated data collection.

\subsection{Representative Photodiode Data}

\begin{figure}[htbp]
\begin{center}
\plotone{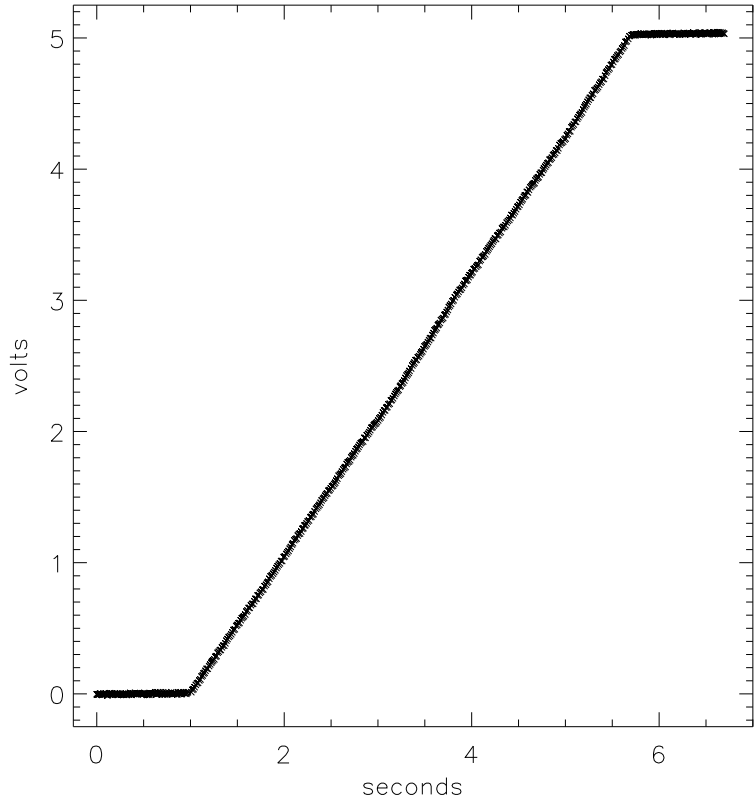}
\caption{Typical integrated calibration diode photocurrent vs.~time. 
The vertical axis is the integrator output (in Volts)
which is proportional to integrated photoelectrons. The horizontal axis is time in seconds. 
The period of laser activity is the steep portion after $t=1 s$. We define the 
two endpoints as $V_{1}^{on}$ and $V_2^{on}$ for data taken when the 
laser is active, and as $V_1^{off}$, and $V_2^{off}$ in the case when 
the laser is off. The
laser light totally dominates over ambient light levels in the dome, in the example
shown here. 
Furthermore, the slope of the integrated signal is quite constant over the 
period when the laser is on, indicating good pulse-to-pulse intensity stability for the 
laser light. 
The laser shutter was closed when the integrator voltage reached the threshold
of 5.0 V.}
\label{fig:integrator}
\end{center}
\end{figure}

Figure \ref{fig:integrator} shows a typical measurement of the integrated light intensity seen by the photodiode and integrator circuit. The 
plot shows (at an acquisition rate of 100 samples/sec) the ambient and laser light contributions. 

The integrator signal shown in Figure \ref{fig:integrator} indicates that the ambient light level 
was well below the laser light level. The integrator time series was analyzed to 
extract four voltages, corresponding to the start and end points, as well as the 
transition points where the laser shutter was opened and closed. For the analysis 
presented here we used only the start and end values $V_1$ and $V_2$, as described below.

\section{DATA PROCESSING}
\label{sec:processing}

\subsection{Calibration Light Intensity Signal Processing}

We digitized and stored the integrated photocurrent signal before, during and after the 
laser shutter was opened. We determined the start and end integrator 
voltages, $V_{1}^{on}$, $V_2^{on}$, $V_1^{off}$, and $V_2^{off}$ respectively, for the 
laser-on and laser-off images.  The difference $\Delta V=V_2-V_1$ is proportional 
to the total amount of photocurrent produced by the photodiode while the camera
shutter was open. 
Dividing this 
value by the photodiode's quantum efficiency  provides us with a quantity that is proportional to the number of photons delivered to the input pupil. 

\subsection{Determination of System Throughput}

For each wavelength we computed the fraction of incident photons that 
each pixel detected. We subtracted an appropriate laser-off frame from 
each calibration image. This compensates for ambient light in the dome, 
for any dark current (which in our case is negligible) and for any bias 
structure in the array. The resulting 
$\Phi_{laser}(\lambda)=\Phi^{on}(\lambda)-\Phi^{off}(\lambda)$ image was 
normalized to the number of incident calibration photons, to arrive at a relative system 
throughput (in arbitrary units that are proportional to 
ADU's per pixel per incident photon, for pixel $i,j$) of

$$T(i,j,\lambda)=\Phi_{laser}(i,j,\lambda)QE(\lambda)/(\Delta V^{on} - \Delta V^{off}),$$

\noindent
where $QE(\lambda)$ is the calibration photodiode's quantum efficiency. 
The $T(i,j,\lambda)$ data cubes for each filter comprise the system sensitivity data 
we seek. 
We stress that this produces a set of efficiency curves for each filter that have a 
common basis, namely the known detection efficiency of the calibration photodiode. 

\subsection{Spatial Averaging}

We have obtained calibrated throughput data for 1.4 billion pixels at over 350 wavelengths. 
In order to reduce this data cube to tractable subsets suitable for visualization 
and human digestion, we have chosen to 
produce averages over 7 annular regions of the $T(i,j,\lambda)$ frames, 
with equal $dr$ values. 

\section{RESULTS}
\label{sec:results}

The data collected here can be used in a variety of ways. 
For this presentation of initial calibration results, we chose to investigate:

\begin{enumerate}

\item{} Short term repeatability of the measurements.

\item{} A determination of relative system throughput vs.~wavelength, tied to a 
common standard: the response curve of the calibrated photodiode.

\item{} The absolute transmission function of each filter, 
by comparing the response function with and without a filter in the beam. 

\item{} Spatial variation in transmission vs.~wavelength, radially across the focal 
plane.   

\item{} A determination of fringing amplitudes vs.~wavelength, for the deep depletion 
CCDs in the PanSTARRS camera.

\end{enumerate}

These are treated in turn, in the subsections that follow.

\subsection{Short Term Repeatability}

For a few wavelengths in the $r$ band we made a succession of 
five measurements, spread over the 48 hour period of data acquisition.
The means and standard deviations of these repeated data points are 
presented in Table \ref{tab:repeats}. These figures are for a single
annular region on the focal plane. It appears that short term stability
is at the level of one to two parts per thousand. 

\begin{table}[htdp]
\caption{Short term Repeatability for 5 Measurements Taken Over
48 hours.}
\begin{center}
\begin{tabular}{cccc}
\hline
\hline
$\lambda$ & Mean T & $\sigma_T$ \\
(nm) & (relative to peak) & ~ \\
\hline
  550   &   0.60 & 0.0016 \\ 
   554  &   0.81&  0.0018 \\
   558   &   0.88 & 0.0018 \\
   562    &  0.90 & 0.0017 \\
   566     & 0.91 & 0.0063 \\
   570     & 0.90 & 0.0013 \\
   574    &  0.89 & 0.0012 \\
   578    &  0.92 &  0.0012 \\
   582    &   0.94 & 0.0019 \\
   586    &   0.93 & 0.0013 \\
   590    &   0.94 & 0.0016 \\
   594    &   0.94 & 0.0022 \\
   598    &    0.96&  0.0015 \\
\hline
\end{tabular}
\end{center}
\label{tab:repeats}
\end{table}%

\clearpage
\subsection{Relative System Throughput Across the PanSTARRS Passbands}

\begin{figure}[htbp]
\begin{center}
\epsscale{1.0}
\plotone{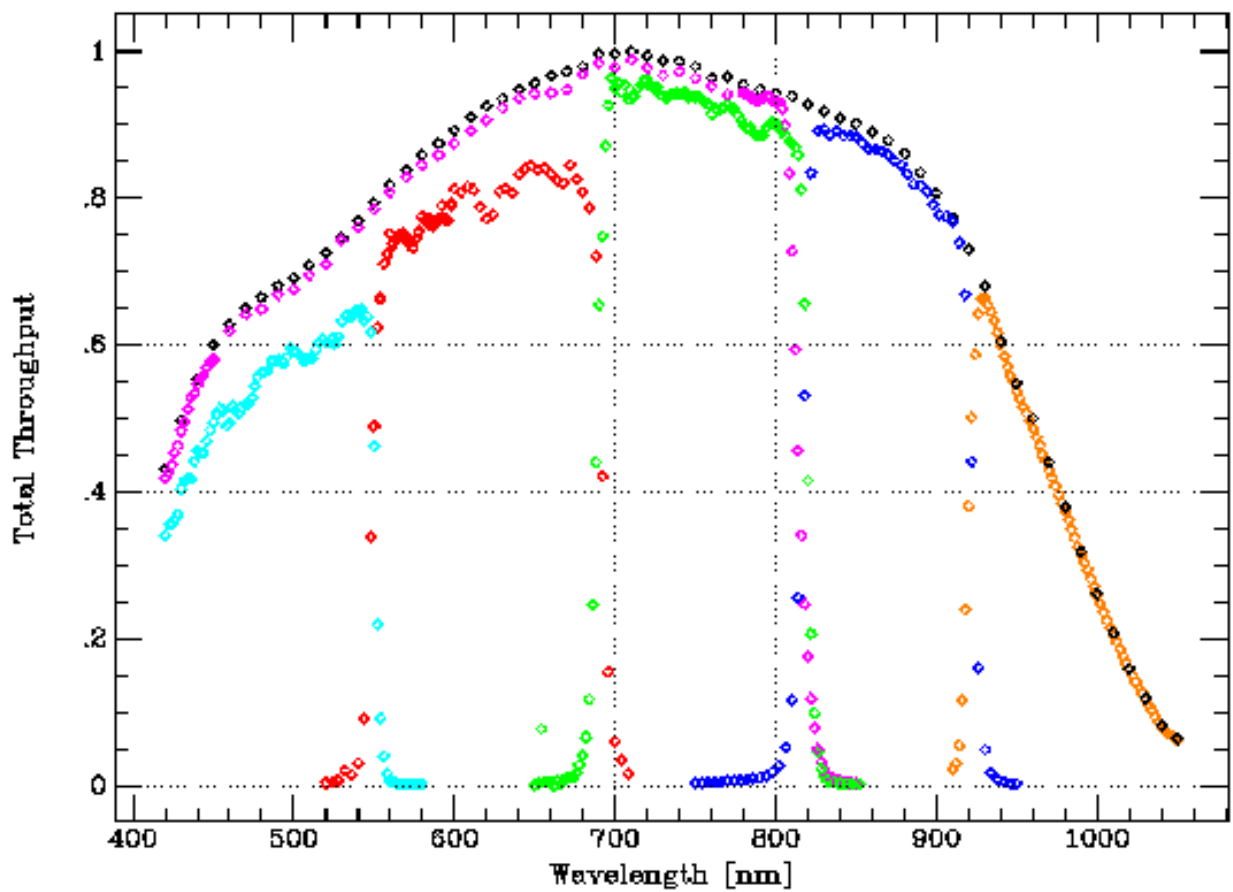}
\caption{PanSTARRS transmission functions, for g (cyan), r (red), i (green), z (blue),
y (orange), w (purple) and open (black), determined from full-pupil illumination 
with tunable laser light. The region between 550 and 600 nm 
was scanned five times, and shows short term repeatability at the level of a few parts per 
thousand.  Since there is a single overall multiplicative free parameter, 
we have chosen for this plot to 
normalize all the curves to the peak sensitivity seen, with no filter in the beam.  }
\label{fig:transmission}
\end{center}
\end{figure}

We have mapped out the integrated system response for the PanSTARRS optics
and camera, with a metrology scheme that allows us to compare the 
throughput across all passbands. 
In Figure \ref{fig:transmission} we show the system throughput curves we
measured. 

Taken in conjunction with an estimate of atmospheric transmission we can use
these curves to compare the {\it colors} from synthetic photometry of 
spectrophotometric standards to the colors obtained from the PanSTARRS
system. 

\subsection{Filter Transmission Functions}

\begin{figure}[htbp]
\begin{center}
\epsscale{1.0}
\plotone{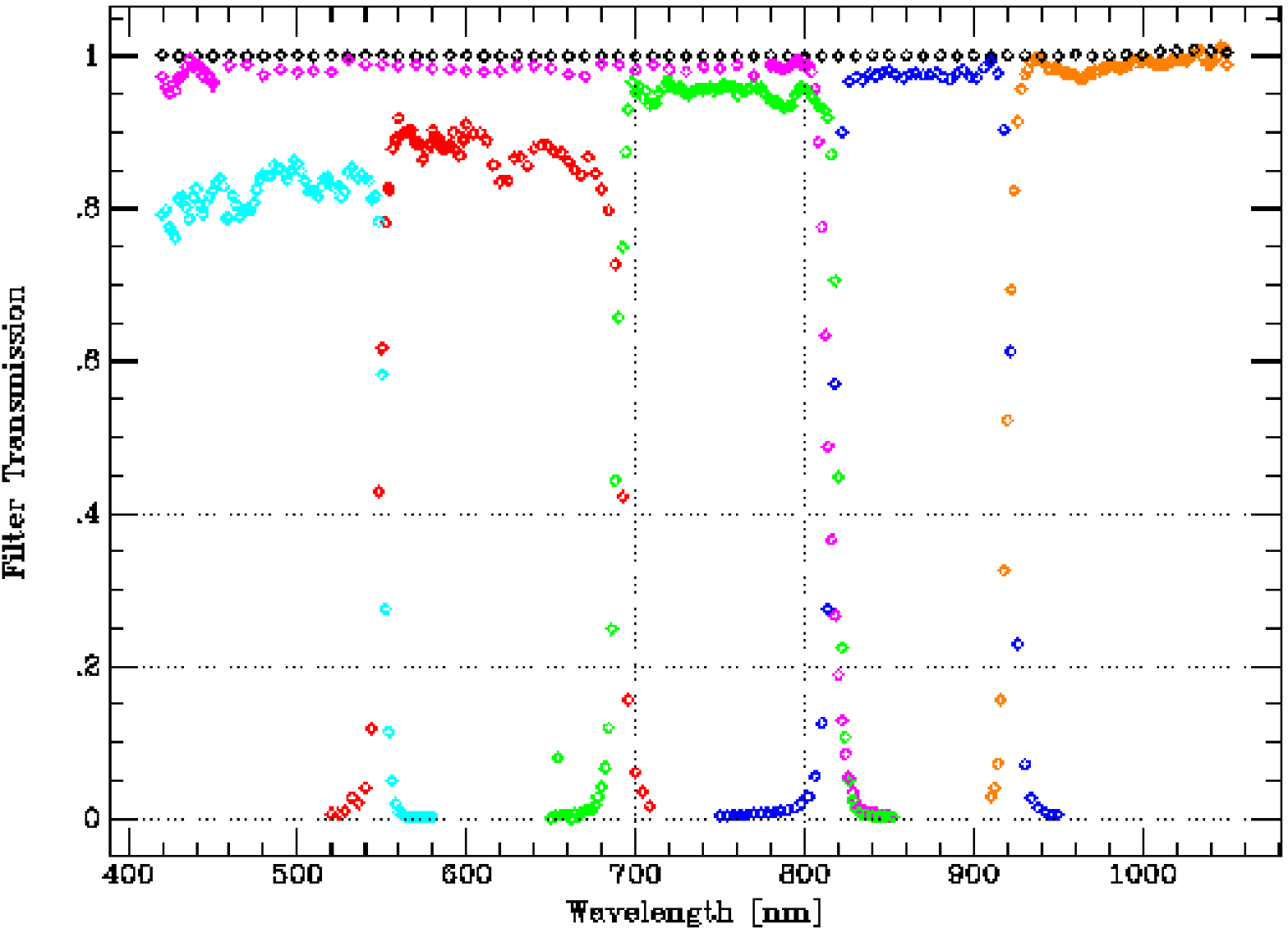}
\caption{PanSTARRS filter-only functions, for g (cyan), r (red), i (green), z (blue),
y (orange), w (purple) and open (black). For this plot the vertical axis does
correspond to the absolute filter transmission, since these data are the ratio
between the measured sensitivity functions with and without the respective 
filters in the beam.}
\label{fig:filters}
\end{center}
\end{figure}

By dividing the transmission curve obtained with a filter in the beam to that obtained
with no filter, we can determine the absolute filter transmission curves with the 
same angular distribution as light from celestial sources. The filter transmission functions we
obtained are shown in Figure \ref{fig:filters}, for one annular region in the focal plane

\begin{figure}[htbp]
\begin{center}
\epsscale{1.0}
\plotone{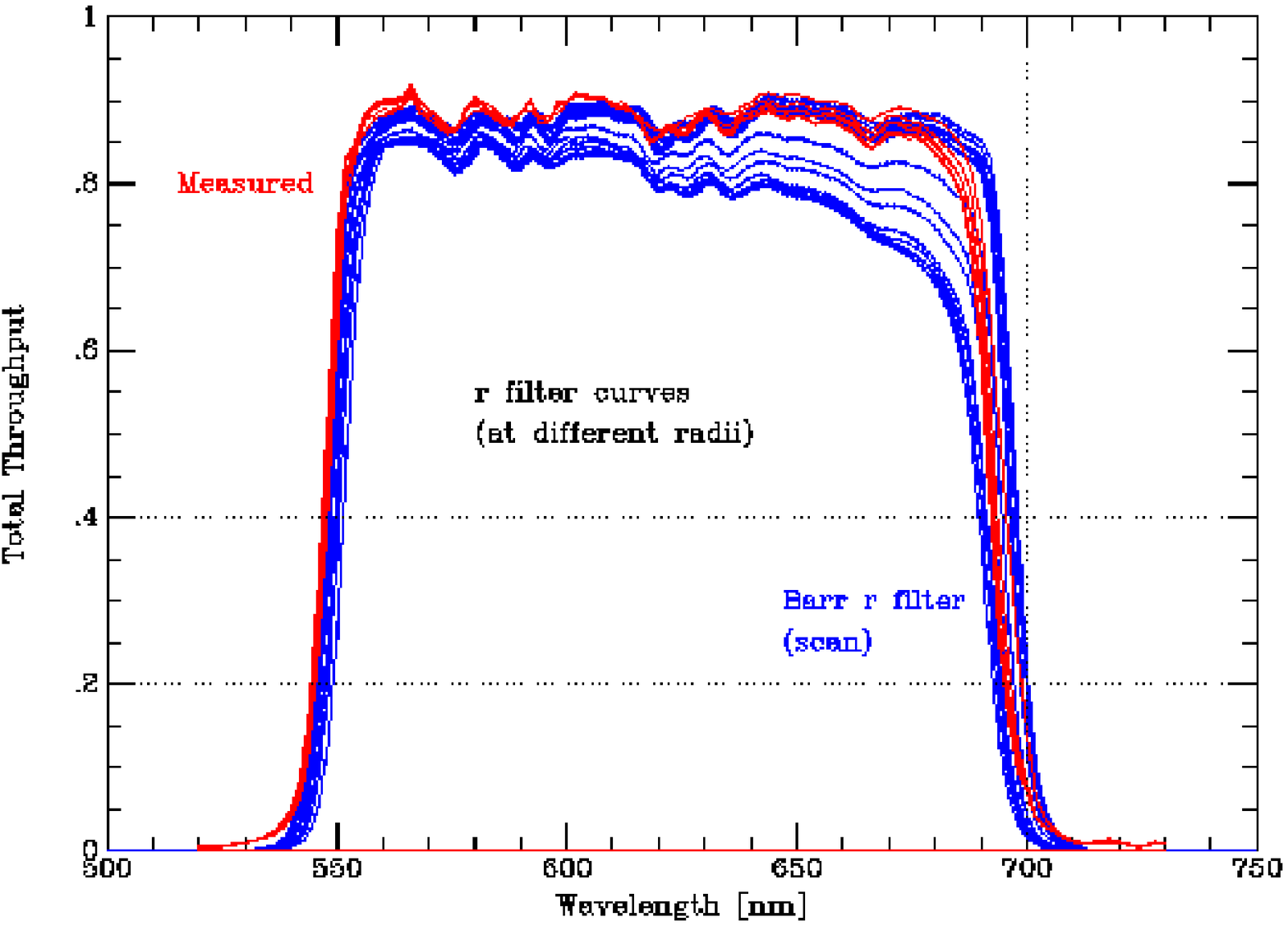}
\caption{
This plot compares the radial variation at the focal plane we observed (red) in
  the Pan-STARRS $r$ filter with the spatial variation in filter transmission reported by the
  vendor, in collimated light (blue).  The vendor curves rise on the
  red end from 1 to 9 inch radius, and the highest cluster of eight
  curves are at radius 9.5 inch, on the eight octagonal faces.
  The microstructure we measure corresponds in detail to that reported
  by the vendor. The radial dependence at the focal plane (what we measure) 
  is expected to be the convolution of the beam footprint at the filter with the filter's 
  spatial variation in transmission (what the vendor measured), with some perturbation due to the angular distribution of converging rays. The data 
  obtained on the telescope exhibit the shift towards the blue that is characteristic 
  of interference filters for light at other than normal incidence. 
  These complications add motivation to making an integral, 
  full-aperture determination of system throughput, as described in this paper.}
\label{fig:rband}
\end{center}
\end{figure}

Figure \ref{fig:rband} compares the benchtop trace of the $r$ band interference
filter \citep{Barr} to what we obtained.  
The microstructure in the main passband is reproduced, and the skirts of the filters
show good correspondence. The family of curves we obtained from different 
annuli in the focal plane shows the same structure in transmission vs.~wavelength.

\subsection{Spatial Variation in Spectral Response}

\begin{figure}[htbp]
\begin{center}
\epsscale{1.0}
\plotone{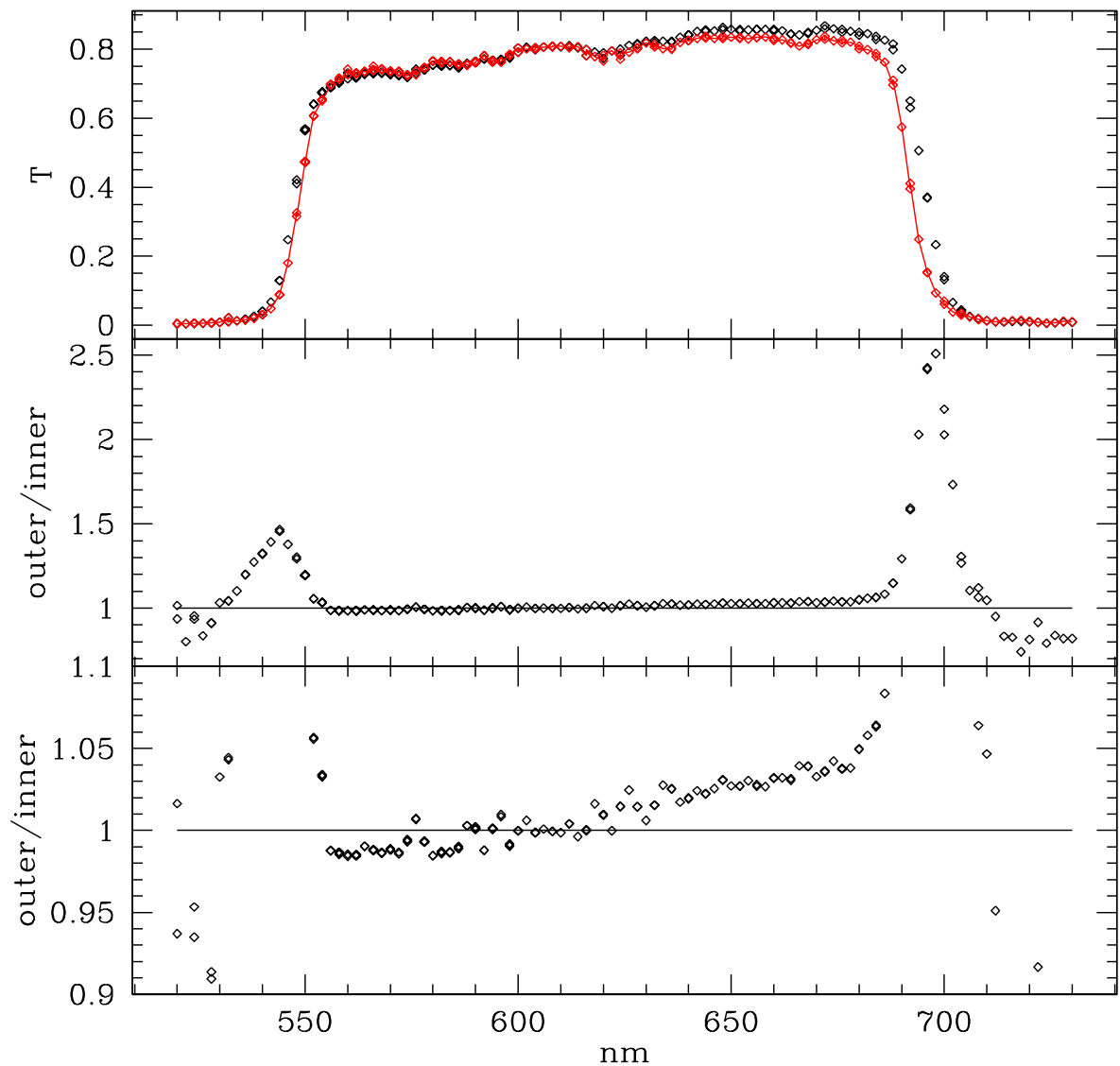}
\caption{The upper panel shows the variation in system response in the $r$ band
between the center (red line) and the outer annulus (black points)  of the PanSTARRS focal plane. 
The middle panel shows the ratio $R= outer/inner$ of the system response function, and the lowest panel 
shows this same ratio at a different scale. There is a 
shift in the effective passband between the center and the edge of the 
field, with the outer  
edge having both an enhanced red response and a wider passband.}
\label{fig:ratioplot}
\end{center}
\end{figure}

We explored the variation in response across the focal plane. This is most 
pronounced at the edges of the filters. Figure \ref{fig:ratioplot} shows the field 
dependence of the $r$ band response of the system. (We see similar 
radial variations in system response at the filter edges in other passbands
as well.)  By taking annular averages in system response to produce Figures 7 and 8, 
we are averaging over any first order gradient in screen uniformity, as discussed
in the description of potential sources of systematic error, below. 

There is clear evidence for non-trivial response variations between the center
and the edge of the field, in the $r$ band. The outer regions of the field have
an $r$-band response that cuts on at a bluer wavelength, and extends to longer
wavelengths before cutting off. In addition, there is an overall enhancement 
in response at longer wavelengths, at the edge of the field . The outer annulus of the instrument
appears to have an $r$ band that is a few nm wider, and a response to wavelengths
$\lambda > $600 nm that is a few percent higher, than the response measured at the center
of the field. 

A certain amount of this field dependence in the effective passband will be 
compensated in the process of constructing an illumination correction, 
{\it i.e.}~by insisting that the natural system magnitudes of sources are independent
of position on the focal plane. However the data we show here indicate that we
should expect a color-dependent residual. 

\subsection{Fringing}

\begin{figure}[htbp]
\begin{center}
\epsscale{0.8}
\plotone{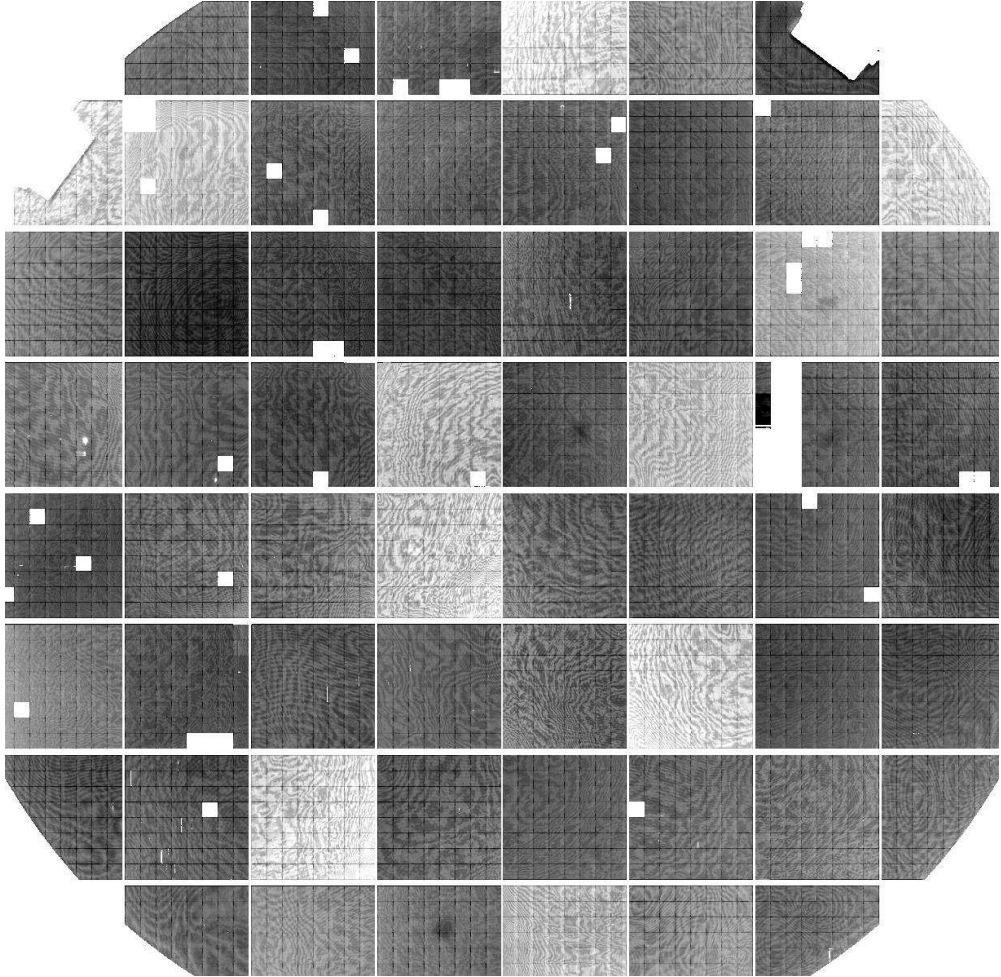}
\caption{Fringing pattern in monochromatic light at 902 nm, in the $z$ band, 
across the PanSTARRS array. The peak-to-peak fringing amplitude 
at this wavelength is 5\%.  (The individual gains of the detectors have not
been normalized, and that accounts for the chip-dependent variations.)}
\label{fig:fringing}
\end{center}
\end{figure}

The deep depletion detectors used in the PanSTARRS camera have significant
quantum efficiency in the near infrared (as seen in the ``open'' curve
in Figure \ref{fig:transmission}). This suppresses fringing compared to 
currently conventional CCDs. Figure \ref{fig:fringing} shows a full frame obtained
in the $z$ band at $\lambda$=902 nm. 

In conjunction with a spectroscopic measurement of the relative intensity of 
OH night sky lines, the images we have obtained could in principle be used
to construct a spectrally-matched fringe frame for each image taken in the 
PanSTARRS survey. Even better would be to obtain monochromatic dome
flats at the specific wavelengths that correspond to the brightest sky lines, 
and then make an flux-weighted sum of those images to construct a fringe frame
appropriate for each PanSTARRS survey image. 

\clearpage
\section{POTENTIAL SOURCES OF SYSTEMATIC ERROR}
\label{sec:systematics}

We present below our current
level of understanding of various sources of systematic error, steps taken to date, 
and what we perceive as the path forward. 
Table \ref{tab:systematics} summarizes 
our current estimate of the impact of systematic errors. 
The subsections following the Table provide the rationale for the 
entries. 
Since we have yet to measure and correct for stray and scattered light effects, 
specifically the difference between system sensitivity to diffuse light as opposed
to resolved sources at the arcsec scale, our systematic error budget
is currently dominated by our conservative estimate of the illumination correction, 
at the 5\% level.  If we succeed in achieving a factor of ten reduction in this
systematic error (which as outlined below has been achieved by others) 
then we might hope to attain $\sim0.2-0.5\%$ fractional uncertainty in 
instrumental throughput determination.
 
\begin{table}[htdp]
\caption{Systematic Uncertainty Estimates for Night-time Throughput Calibrations. }
\begin{center}
\begin{tabular}{lr}
\hline
\hline
~~~~~~Gremlin & Estimated $\sigma_T/T$~~~~~ \\
\hline
Photodiode Calibration Uncertainty & $10^{-3}$ for 470 nm $< \lambda < $ 950 nm\\
~ & 1.0 to 1.5 $\times 10^{-2}$ for $\lambda > $950 nm\\
~&  5.0 to 1.0 $\times 10^{-3}$ for 350~nm $< \lambda < $470 nm.\\
 Ambient Light Fluctuations & $<2 \times 10 ^{-3}$ \\
 Stray Light,  ``Illumination'' Corrections$^a$&  $5 \times 10^{-2}$ \\
 Wavelength Independent Screen Non-Uniformity & $5 \times 10^{-2}$\\
 Wavelength Dependent Screen Non-Uniformity$^b$ & 2\%\\
 Spectral Purity & $<5\times10^{-4}$ for $\lambda>$400 nm\\
 Wavelength Calibration & $10^{-2}$ on steep filter skirts \\
 ~&$<10^{-3}$ otherwise\\
 Timing Latencies & $<10^{-4}$ \\
Camera Shutter Artifacts & $< 10^{-4}$\\
CCD or Photodiode Temperature & 1\% per degree C for $\lambda > 900 $nm\\
\hline
\end{tabular}
\end{center}
$^a$This is our conservative estimate, with no corrections applied.
\cite{marshall} have demonstrated $6\times10^{-3}$.

$^b$ This applies to the $r$ band, currently unconstrained at other wavelengths.  
\label{tab:systematics}
\end{table}%

\subsection{Uncertainty in the Photodiode Calibration}

Our throughput measurement is
only as good as the comparison standard we use. 
The combination of the QE curve of the photodiode, the photodiode's optical 
train, and the integrator circuit's capacitor comprise the metrology standard
against which we are determining total system throughput. Fractional calibration 
uncertainty in the reference diode's QE curve is reported \citep{diode, NIST} by NIST as 0.2\%
at 400 nm, 0.1\% between 470 and 950 nm, rising 
to 1\% for $\lambda > 950$nm.\footnote{NIST's convention is to quote 
2$\sigma$ uncertainties ($k=2$) and we have adjusted this to equivalent 1$\sigma$
values, assuming Gaussian statistics.} 
This systematic uncertainty in QE dominates over other considerations such as 
the temperature coefficient of the integration capacitor times temperature
changes over the duration of the measurement. 

Even if dust flecks land on the calibration diode window in the time between 
the NIST calibration and our measurement, 
as long as this is a ``grey'' reduction in sensitivity it does
not affect our relative throughput measurement.  However monitoring the 
overall system throughput over extended periods of time will rely on both the optical configuration and 
diode sensitivity remaining constant.

\subsection{Ambient Light Fluctuations}

Our calibration technique relies upon our being able to determine the system's response
to the light from the Ekspla NT242 OPO, and we do this by subtracting (from both the 
Gigapixel images and the photodiode signal) the flux seen in an 
equal amount of time.  If the ambient background light comprises a 
fraction $f$ of the {\it detected} flux in the imager, and this background varies by a 
fraction $\delta$ in the time between the illuminated and the background image, 
a 1\% measurement of system sensitivity requires $f\delta < 0.01$. 
Particularly on the skirts of the filters, where $f$ increases 
(since the instrument is detecting background light 
across the filter's entire optical passband, but sensitivity to the 
calibration light is suppressed), this strongly favors taking the calibration
images on cloudy nights, when the ambient light level is both low and stable. 

For the images taken at night, background light in the dome is essentially
undetectable. A conservative estimate of the impact of ambient light fluctuations
comes from the results shown in Table \ref{tab:repeats}, which includes
daylight measurements, at $\sigma_T /T < 2 \times 10^{-3}$. 

\subsection{Stray and Scattered Light Paths}

Stray light is a potential problem
when dome flats are used for sensitivity corrections, and
our approach is not immune to this concern. 
Non-focused  
light paths can add flux to that from the 
direct, focused, path through the optical train. Only the wavelength-dependent
variation in the stray light distribution will produce a 
systematic error in our throughput determination. Wavelength-dependent ghosting 
in the optics  is one example of a scattered light path that might
produce a variable gradient in illumination across the focal plane. This would then be 
misinterpreted as sensitivity variation, even if the system sensitivity is uniform. 
It is therefore important to obtain ``illumination corrections'' using point sources
on the sky. We have not yet made this comparison for the PanSTARRS imager, 
but experience with other telescopes suggests that the illumination correction
could be as large as 5\%, and we will adopt this as our current estimate 
of the systematic error for stray and scattered light. We consider this to be
a very conservative estimate, since only the wavelength-dependent portion of the 
illumination correction will afflict our determination of system throughput. 

\cite{marshall} have demonstrated using dome flats to achieve consistent 
photometric zeropoints at a level ten times better than our conservative
estimate of systematic errors due to our not having made illumination 
corrections. 
\cite{SNLS09} describe millimagnitude residual uncertainties after 
making careful illumination corrections, and in the future  
we hope to achieve similar performance, after measuring and 
making illumination corrections. 

An illumination screen that only projects light over the range of angles seen
by the camera ($\sim$3$\cwsdeg$ for PanSTARRS and LSST) would greatly reduce
the stray light contribution. We have such a system currently under development 
\citep{LSST}. 

\subsection{Wavelength-Independent Illumination Screen Non-Uniformity}

The camera pixels sense light primarily from a common area on the 
flatfield screen. We estimate that a 10\% fractional uniformity in surface brightness 
should produce illumination on the focal plane that is uniform at the 1\% level. 
We did not achieve this level of uniformity in the configuration used
to obtain the data described here, so there is a potential grey multiplicative 
variation across the focal plane. 

As long as the spatial surface brightness distribution of the screen does not
exhibit a significant variation with wavelength, we can reliably use the data to 
assess the spectral sensitivity of each pixel, independently. Wavelength independent
illumination variations impart a multiplicative grey variation in measured system 
response across the focal plane. This then falls into the category of effects that 
will compensated for with an on-sky illumination correction, and the systematic
error from wavelength-independent screen illumination is degenerate with other
illumination correction effects, such as vignetting, etc. We therefore adopt the same
systematic uncertainty here as that for stray and scattered light, $5 \times 10^{-2}$.

\subsection{Wavelength-Dependent Illumination Screen Non-Uniformity}

Comparing pixel response functions across
the focal plane is undermined by any wavelength-dependent variations in focal plane
illumination. Furthermore, 
if the pixels in the camera and the calibration photodiode intercept a wavelength-dependent difference in 
flux from the screen, a systematic error is introduced in the wavelength 
response of even a single pixel. Finally, even if the screen illumination is completely flat across all wavelengths, spectral variation in the reflections that 
feed stray and scattered light
paths could introduce a systematic error in the measured response function. 

A careful and thorough investigation of these different contributions will be 
a topic for future work, and for the purpose of this paper we will restrict our
attention to the $r$ band. 

The data in 
Figures 7 and 8 indicate that away from the filter skirts, annular averages of 
$r$ band system response (which eliminate first order gradients across the focal plane)
are radius-independent to a few percent. 
Pending further analysis we will adopt the value of 2\% as an estimate of the 
potential systematic error due to wavelength-dependent illumination effects, 
within the $r$ band, but further investigation is needed. 

\subsection{Spectral Purity and Optical Bandwidth}

The tunable Ekspla NT242 OPO produces light at numerous wavelengths (355, 512, 1064 nm) 
in the process of generating the photon pair from the downconversion in the OPO. 
Of particular concern would be leakage of one of these unwanted wavelengths
in the passband being measured, or leakage light conspiring with (say) a red
leak in an interference filter. We used the Ocean Optics spectrograph to verify
that (with one exception) the light sent into the optical fiber had undetectably
low ($<5\times10^{-4}$) contamination light. The exception arises when wavelengths 
$\lambda<$~400~nm 
are requested, which the Ekspla NT242 OPO produces by frequency-doubling 
of OPO output light at 2$\lambda$.  In this regime there is significant leakage
light, which led us to avoid observations at the blue edge of  the $g$ band, pending
the installation of a short-pass filter. While the $g$ filter would likely sufficiently block the 
red leakage light from reaching the instrument, the calibration diode would see
this light from the screen, producing a systematic error in the reported flux. 
In a subsequent run will place a short-pass filter in the Ekspla NT242 OPO for scanning 
the region below 400 nm. 

The optical bandwidth of the calibration light is another potential source of 
systematic error, since the measured response function is the actual 
monochromatic response convolved with the bandwidth of the calibration light. 
At the longest wavelengths
we probed, $\lambda\sim1\mu$m, the calibration light had a bandwidth of under 0.5 nm.
This consideration is one of the main reasons we favor the use of the tunable 
laser over a monochromator, from which it is difficult to achieve the 
desired spectral power density. 

\subsection{Wavelength Calibration}

An error in wavelength calibration will of course shift the measured curves 
accordingly. We showed above that our wavelengths are known to an uncertainty
of $\sigma_{\lambda}\sim0.2 nm$. This effect is most pronounced on the steep 
edges of the interference passbands, and the error introduced in transmission 
is $\sigma_T=\sigma_{\lambda} dT/d\lambda = 0.2 \times 0.06 \sim 1\%$, since the steepest
slope we see in the filter skirts is $dT/d\lambda = 0.06~$nm$^{-1}$. Away from the filter skirts
the systematic error due to wavelength miscalibration is a factor of ten lower, under a part per thousand. 

\subsection{Timing Latencies and Shutter Artifacts}

Although the PanSTARRS  image headers do contain actual shutter open time 
with msec accuracy, our mode of operation is inherently insensitive to timing jitter.
Our method of taking data assures that each pixel is exposed to the calibration 
light for exactly the same time period, regardless of shutter imperfections or
asymmetries.
As long as the fixed dose of laser light is provided 
during the interval
we are certain the camera shutter is open, variation in shutter open times
only influences our correction for ambient light levels, and at night this
is negligible. We therefore assess the systematic error from shutter open
timing to contribute less than $10^{-4}$ to our flux uncertainties, for night-time 
calibration data.

Another potential source of background light mis-correction comes from 
network latency in the client-server configuration we used. 
We will adopt an estimate of 5 msec for jitter in the timing
between the computers, but again this only influences the 
ambient light correction, which for night-time calibration images
is negligible anyway,  and so we assign this a contribution 
of $< 10^{-4}$. 
 
\subsection{CCD and Photodiode Temperature Variations}

The process of converting incident photons
into photoelectrons is assisted by thermal excitation of electrons into the 
conduction band of Silicon. The Gigapixel camera's sensitivity is therefore
a function of the focal plane temperature. This is most pronounced in the $y$ band, 
especially since the red edge of the $y$ band is defined by the falling 
CCD QE curve. Similarly, the sensitivity of the calibration photodiode to NIR
light depends to some extent on its temperature. 

As long as both of these dependencies are known, we can account for the
temperature-dependence of the sensitivity of both the calibration diode and the 
focal plane array. We can measure the temperature of the calibration diode with 
sufficient accuracy to properly account for its QE(T,$\lambda$). The thermometry 
on the Gigapixel camera is somewhat more problematic, since the temperature sensors
are some distance from the detectors. Our approach to this in the future will be to map 
out the system response function at a variety of camera temperature settings, and 
establish an empirical relationship between system throughput and the temperature
data provided by the camera. We can then use an appropriate response function 
for each survey image, based on the temperatures reported in the image headers. 

For the data presented here, we assess the systematic due to temperature-dependent QE effects to be under $10^{-3}$ for $\lambda<900~nm$, rising to 1\%/$\cwsdeg$C
at 1 $\mu$m. 

%\clearpage
\section{DISCUSSION}
\label{sec:discussion}

We have shown the ability to obtain a succession of monochromatic 
calibration images that exhibit short term repeatability at the level of a few
parts per thousand. 
Our long term goal is to provide, for each survey image and 
associated photometric catalog, the effective passband through which the image
was acquired. This would include both the instrumental response function 
at the location of the source, as well as the estimate of atmospheric transmission
along the appropriate line of sight. 

These results encourage us to further pursue this
technique to both measure and monitor system sensitivity over the course of the 
PanSTARRS survey.
We also intend to use the system described here to measure  
the dependence of the response function on CCD device temperature, in the 
phonon-assisted photoconversion regime at wavelengths near 1 $\mu$m. 
Another interesting measurement would be to compare the response function 
before and after mirror cleaning or dust deposition events. 

We can also use these instrumental sensitivity data in conjunction with 
measurements of atmospheric 
transmission to compare synthetic photometry with observations, as described in 
\cite{Calib06}, across different regions of the focal plane. Table \ref{tab:sowhat}
is an initial exploration of this approach. We took model atmospheres from 
\cite{Kurucz96} and converted them to photon spectral energy (PSED) distributions
$\Phi_\gamma(\lambda)$. We then
integrated\footnote{We have omitted the effect of atmospheric attenuation since our
interest here is only to explore the effect of focal plane position-dependence
of the sensitivity function.} 
these photon distributions over the seven distinct annular $r$ 
band response functions, to assess the photometric perturbation that would
arise from the observed radial variation in the sensitivity functions. 
We computed for each radial annulus $a$ the ``raw'' synthetic magnitude
$$ m(a)_{raw}=\int T(a,\lambda) \Phi_\gamma(\lambda) d\lambda,$$
and also a normalized magnitude that corresponds to constructing a 
photon-flat 
$$ 
m(a)_{normalized}= \frac{\int T(a,\lambda) \Phi_\gamma(\lambda) d\lambda}
{\int T(a,\lambda)  d\lambda},
$$ where $a$=1,2...7. The normalized magnitudes correct for the radial 
variation in the integrated area under the sensitivity curve. 

Table \ref{tab:sowhat} shows the 
resulting magnitude differences, $\Delta m=m_{edge}-m_{center}$, between the
edge ($a$=7) and the center ($a$=1) of the PanSTARRS field, for different stellar types. 
The second column in Table \ref{tab:sowhat} shows that the raw flux at the field edge 
ranges  from  3.7\% to 4.6\% higher than the value at the center, with the variation 
being monotonic with decreasing stellar temperature. 
This is consistent with the sensitivity function behavior shown in Figure~\ref{fig:ratioplot}: stars at the edge appear brighter, and red stars more so than blue ones.

If we compare instead the 
``normalized'' magnitudes (the third column in Table \ref{tab:sowhat})
the flux excess diminishes by a factor of 5 to 20, depending on stellar 
temperature. However, even after flux normalization at each radial annulus by   
$\int T(a,\lambda)  d\lambda$, a scalar quantity that implicitly weights each 
wavelength equally,
the third column in the Table still shows up to 
$\sim$1\% perturbations in stellar photometry ({\it e.g.}
 $\Delta$m$_{normalized}$(O5V) -  $\Delta$m$_{normalized}$(M6V)=0.0094) that depend on 
both stellar temperature and separation across the field.  

\begin{table}[htdp]
\caption{Influence of Observed $r$ band Center-to-Edge Variations on Synthetic Stellar Photometry. }
\begin{center}
\begin{tabular}{ccc}
\hline
\hline
Stellar & Edge-Center & Edge-Center \\
Type & ~~~$\Delta$m$_{raw}$~~~& $\Delta$m$_{normalized}$ \\

\hline
 O5V  & -0.0367 & ~0.0068 \\ 
 O7V  & -0.0368 & ~0.0068 \\ 
 O9V  & -0.0368 & ~0.0067 \\ 
 B1V  & -0.0371 & ~0.0065 \\ 
 B3V  & -0.0374 & ~0.0061 \\ 
 B5V  & -0.0376 & ~0.0059 \\ 
 B8V  & -0.0379 & ~0.0056 \\ 
 A1V  & -0.0385 & ~0.0050 \\ 
 A3V  & -0.0388 & ~0.0047 \\ 
 A5V  & -0.0393 & ~0.0042 \\ 
 F0I  & -0.0391 & ~0.0044 \\ 
 F0V  & -0.0391 & ~0.0044 \\ 
 F2V  & -0.0411 & ~0.0025 \\ 
 F5V  & -0.0417 & ~0.0019 \\ 
 F8V  & -0.0420 & ~0.0015 \\ 
 G2V  & -0.0425 & ~0.0011 \\ 
 G5I  & -0.0425 & ~0.0011 \\ 
 G5V  & -0.0426 & ~0.0009 \\ 
 G8V  & -0.0429 & ~0.0007 \\ 
 K0V  & -0.0434 & ~0.0001 \\ 
 K4V  & -0.0450 & -0.0015 \\ 
 K7V  & -0.0460 & -0.0025 \\ 
 M2V  & -0.0457 & -0.0021 \\ 
 M4V  & -0.0461 & -0.0026 \\ 
 M6V  & -0.0461 & -0.0026 \\ 
  \hline
\end{tabular}
\end{center}
\label{tab:sowhat}
\end{table}%
 
We conclude from this synthetic photometry exercise that we can expect 
half-percent level $r$ band zeropoint shifts between the center and the edge of the field,
even after flat-fielding, for a given stellar type. Looking across stellar types 
(from O to M class stars) we can expect systematic zeropoint differences 
at the percent level, between the center and the edge of the field.   As 
outlined in \cite{Calib06}, knowing the system throughput for each pixel 
and for each wavelength will allow us to account for these effects in detail. 

This suggests that pushing the photometric precision beyond the 1\% level 
will likely require a more sophisticated flat-fielding approach, that takes full 
advantage of the system's measured spectral response function
(as well as a determination of atmospheric transmission).  Also, the determination 
of ``illumination corrections'', by rastering a source around the focal plane
and requiring consistent results, will be PSED-dependent at the percent level. 

We are now in a position to quantitatively assess various flat-fielding schemes, such as  
performing a joint analysis of photometric data from all bands, that take into 
account the position-dependence (and also, if needed, time-dependence) of the 
system's response function. 
 
Finally we note that a full-aperture calibration taken of a distant source of
known radiance, even at just one wavelength, would suffice to 
determine the single overall multiplicative term
we need to extend these measurements to an absolute calibration of the 
PanSTARRS system.  This would require appropriate knowledge and correction
for atmospheric transmission, but the single wavelength could be judiciously chosen
(such as $\lambda \sim 808$~nm, also convenient for laser diode sources) to coincide with 
good detector QE as well as high and stable atmospheric transmission.

{\it Facilities:}
 \facility{PanSTARRS}

\acknowledgments

We are grateful to the National Institute for Standard and Technology (under 
award 70NANB8H8007), the LSST Corporation, Harvard University and the Department of Energy Office of Science (under grant DE-FG02-91ER40654) for their support of this work. The dedication and competence of the scientific and technical staff of PanSTARRS project were essential to the success we report here. In particular we are
grateful to Will Burgett, Robert Calder, Ken Chambers, Greg Gates, 
Tom Melsheimer, Jeffrey Morgan, and Shannon Waters for their invaluable assistance.
Conversations with Tim Axelrod, David Burke, Darren DePoy, Ned Henry, David Hogg, 
Paul Horowitz, Zeljko Ivezic, Eli Margalith, 
John McGraw, Armin Rest, Abi Saha,  Nick Suntzeff, Chris Smith, 
Will High and Pete Zimmer were very valuable in designing and refining the technique
described here. We are also grateful to the anonymous hitchhikers in the 
Haleakala National Park who provided us with the opportunity to accumulate 
the good karma needed for this endeavor.

\end{document}